\documentclass[epj]{svjour}
\usepackage{graphics}
\begin{document}
\newcommand{\dE}   {\ensuremath{\mathrm{\Delta E}}}
\newcommand{\Mbc}  {\ensuremath{\mathrm{M_{bc}}}}

\newcommand{\OmC}  {\ensuremath{\mathrm{\Omega_C^0}}}
\newcommand{\JPsi} {\ensuremath{\mathrm{J/\Psi}}}
\newcommand{\epem} {\ensuremath{\mathrm{e^+e^-}}}
\newcommand{\Ys}   {\ensuremath{\Upsilon(\mathrm{4S})}}
\newcommand{\DSP}  {\ensuremath{\mathrm{D^{*+}}}}
\newcommand{\DSpm} {\ensuremath{\mathrm{D^{*\pm}}}}
\newcommand{\DS}   {\ensuremath{\mathrm{D^*}}}
\newcommand{\DP}   {\ensuremath{\mathrm{D^+}}}
\newcommand{\DZ}   {\ensuremath{\mathrm{D^0}}}
\newcommand{\KM}   {\ensuremath{\mathrm{K^-}}}
\newcommand{\PM}   {\ensuremath{\mathrm{\pi^-}}}
\newcommand{\PZ}   {\ensuremath{\mathrm{\pi^0}}}
\newcommand{\PP}   {\ensuremath{\mathrm{\pi^+}}}
\newcommand{\Dj}   {\ensuremath{\mathrm{D_{sJ}}}}
\newcommand{\DsJ}  {\ensuremath{\mathrm{D_{sJ}(2317)}}}
\newcommand{\DssJ} {\ensuremath{\mathrm{D_{sJ}(2457)}}}
\newcommand{\ecS}  {\ensuremath{\mathrm{\eta_c(2S)}}}
\newcommand{\MeV}  {\ensuremath{\mathrm{MeV}}}
\newcommand{\MeVc} {\ensuremath{\mathrm{MeV/c^2}}}
\newcommand{\GeV}  {\ensuremath{\mathrm{GeV}}}
\newcommand{\GeVc} {\ensuremath{\mathrm{GeV/c^2}}}

\newcommand{\ccbar}{\ensuremath{\mathrm{c\bar c}}}
\newcommand{\bbbar}{\ensuremath{\mathrm{b\bar b}}}
\newcommand{\BBbar}{\ensuremath{\mathrm{B\bar B}}}
\newcommand{\frf}  {fragmentation function}
\newcommand{\xP}   {\ensuremath{\mathrm{x_P}}}
\newcommand{\pb}  {\ensuremath{\mathrm{pb^{-1}}}}
\newcommand{\fb}  {\ensuremath{\mathrm{fb^{-1}}}}
\title{Spectroscopy and Charm Quark Fragmentation}
\author{Rolf Seuster% \and Second author\inst{2}% etc
}                     % Do not remove
\offprints{}          % Insert a name or remove this line
\institute{Department of Physics and Astronomy,
University of Hawaii at Manoa, Honolulu, HI 96822
}
\date{Received: \today / Revised version: date}
\abstract{
The large data sample accumulated at the KEKB storage ring allows for
dedicated analysis in charm spectroscopy. This made the first
observations of e.g. the broad $\mathrm{D^{**}}$ resonances in $\mathrm{B}$
decays as well as other processes possible. The observation of the
\DsJ\ resonances are confirmed and indications of their quantum
numbers are given.
The fragmentation function for \DSP\ has been measured with
unprecedented accuracy.
\PACS{
      {13.25.Hw}
      {13.66.Bc}
      {14.40.Lb}
     } % end of PACS codes
} %end of abstract
\maketitle
\section{Introduction}
\label{intro}
Although known as $\mathrm{B}$-factories, the asymmetric storage rings PEP-II
and KEKB enable the BaBar and Belle detectors to explore the charm
sector. Charmed hadrons are not only the dominant decay product of $b$
flavoured mesons, but, in addition, the decay kinematics give valuable
constraints allowing the detection of previously unobserved particles
and the determination of their quantum numbers. The cross section of
\epem\ into charm is similar to the \BBbar\ production cross
section, opening a second door to the charm section via \epem
annihilation.

Due to the excellent performance of the accelerators, the data
sample available for charm studies at Belle is unprecedented. KEKB
recently reached its design luminosity of ${\cal
L}=10^{-34}/\mathrm{cm^2/s}$. Belle accumulated a total integrated
luminosity of about 159~\fb. About 143~\fb\ was recorded at the
\Ys\ resonance, with the largest part of the remaining 16 
\fb\ about 60 MeV\ below the resonance. In the following a few recent
results from the Belle collaboration are presented, see
\cite{BelleWeb} and references therein for more details on the analyses.

\section{Spectroscopy}
\label{sec:1}
\subsection{Charmonium: $\mathrm{\eta_c(2S)}$}
\label{sec:1.1}
Recently, Belle has observed the \ecS\ in $B$ decays at a
mass of $(3656\pm6\pm8)$ \MeVc\, about 60 \MeVc\ higher than the only
previous observation, due to Crystal Ball.
A second process where it was seen is in double \ccbar\ production in
Belle. It
describes the production of a \JPsi\ produced in conjunction with
another \ccbar\ pair in \epem\ annihilation. From non-relativistic QCD
(NRQCD), it is believed that at $\sqrt{s}\approx 10.6 \GeV$ this
process contributes about 10\% to the total $\epem\to\JPsi X$ cross
section. Recently, it was found that double \ccbar\ production is the
dominant contribution \cite{Uglov} and \cite{ccbar}.

Here, a data sample of ${\cal L}=101.8\fb$ has been used.
The \JPsi's are inclusively reconstructed, a
minimum momentum cut effectively removes decay products from $b$ decays.
In the mass spectrum of the system recoiling against the \JPsi, clear
signals of the $\mathrm{\eta_c}$ ($175\pm23$ events) and the
$\mathrm{\chi_{c0}}$ ($61\pm21$ events) are found. Their fitted masses
are in very good agreement with the world averages.
A third peak at a mass of $(3630\pm8)$ \MeVc\ containing $107\pm24$
events (4.4 $\sigma$) is found, identifying it with the \ecS.
The mass determined here is lower than the other determination from
Belle by about $2 \sigma$, but again significantly higher than the one
by Crystal Ball. 

This analysis also confirms the surprising large double charm cross section
over the single charmonium cross section in \epem\ annihilation. More
details about this topic can be found in \cite{Uglov} and \cite{ccbar}.

Now, the \ecS\ has also been seen in two photon events in BaBar and
CLEO.

\subsection{Charmed Baryons: \OmC}
\label{sec:1.2}
The production of baryons in general and of those containing charm in
particular are not well understood. Exact determination of their
properties can help improve understanding these particles. For example,
the mass of the \OmC, a $css$ state and the heaviest weakly decaying
hadron containing 
single charm, is known only on the permille level, with a spread of
about $2.5\sigma$ between the only two different measurements by CLEO
and E687. In a data sample of ${\cal L}=87.1\fb$,
the \OmC\ has been reconstructed in the
$\mathrm{\Omega^-(\Lambda^0K-)\pi^+}$ channel. A momentum cut has been
applied to reduce combinatorial background. The fit to the
reconstructed mass spectrum of candidate events gives $80.5\pm10.8$
observed \OmC\ candidates. From this fit, the mass has been determined
to be $\mathrm{M_\OmC}=(2.693.9\pm1.1\pm1.4)~\MeVc$. This is in good
agreement with the CLEO measurement, but disfavours the E687 measurement.

After extrapolation to the full kinematic momentum range, the
production cross section times branching ratio has been determined to be
$(19.1\pm9.0)$ fb. Here, the largest uncertainty comes from the
shape of the fragmentation function used for the extrapolation.

In addition, the \OmC\ has been reconstructed in the semi-leptonic
decay channel to $\mathrm{\Omega^-\ell^+\nu}$. Interference between
the $s$ quarks from the original quark content and from the decay of
the charm quark can significantly enhance the semi-leptonic branching
ratio. Here an $\mathrm{\Omega}$ is combined with a lepton ($e$ or
$\mu$) of the correct charge to form a \OmC\ candidate. Similar yields
are observed in both channels.
The ratio of branching ratios of the above mentioned hadronic decay to
this semi-leptonic channel has been measured to be $(0.8\pm0.2)$ and is
higher than the previous CLEO measurement of $(0.4\pm0.2)$, but still
consistent given the large uncertainties.

\subsection{D Mesons: $\mathrm{D^{**}}$}
\label{sec:1.3}
In the $c\bar{u}$ system, the large mass of the $c$ quark allows the
application of heavy quark effective theory (HQET).
Assuming an infinite mass for the
$c$ quark, its spin decouples and the total angular momentum $j_q$ of the
light quark becomes a good quantum number. Therefore, four different
P-wave excitations in the $c\bar{u}$ system exist. The two states with
momentum $j_q=3/2$ have already been established. The other two with
momentum $j_q=1/2$ have not been observed yet; a possible
observation by CLEO was never published. They are expected to be
broad, making their observation difficult.

In a sample of 65 million $\mathrm{B\bar{B}}$ mesons, Belle searched
for these two broad states. Decays of charged $\mathrm{B^-}$ to
$\mathrm{D^{(*)+}\pi^-\pi^-}$ have been fully reconstructed. In both
channels, signal yields of $1101\pm 46$ (\DP) and $578\pm 30$ (\DSP)
candidate events were obtained from a fit to the \dE\
distributions over very low background due to the two pions of
negative charge. \dE\ describes the difference of the beam energy and
the energy of the reconstructed $\mathrm{B}$ candidate.

The decay $\mathrm{B^-\to D^+\pi^-\pi^-}$ has previously not been observed.
Here, its branching ratio has been determined to be
$\mathrm{{\cal B}(B^-\to D^+\pi^-\pi^-)=(1.02\pm0.04\pm0.15)\times10^{-3}}$.
The\\ branching ratio of the other decay has been measured to be 
$\mathrm{{\cal B}(B^-\to
D^{*+}\pi^-\pi^-)=(1.25\pm0.08\pm0.22)\times10^{-3}}$. It is lower
than, but within uncertainties compatible with the world average.

An unbinned fit to the Dalitz plot has been performed to
disentangle the different contributions from various intermediate
resonances. The background has been estimated from a fit to sidebands
in \dE.
Apart from the narrow $\mathrm{D_2^{*0}}$ resonance, a
significant contribution of a broad resonance is necessary in order
describe the data. The narrow resonance shows a clear structure of a
tensor particle in the Dalitz plot as expected from the
$\mathrm{D_2^{*0}}$. The broad resonance has never been observed
before and is identified with the $\mathrm{D_0^{*0}}$. Its mass has
been determined to
$\mathrm{M_{D_0^{*0}}}=(2308\pm17\pm15\pm28)~\mathrm{MeV/c^2}$.
The third uncertainty is due to model dependence and is estimated
using 
the differences in fitted masses when neglecting certain particles in
the fit to the Dalitz plot.
The reduced $\chi^2$ of the fit improved when including two virtual
particles, a $\mathrm{D}_v^*$ and a $\mathrm{B}_v^*$, parametrising
contributions from higher resonances.
The same fit gives product branching ratios of the decays of 
$\mathrm{B^-\to D^+\pi^-\pi^-}$ via a $\mathrm{D_2^{*0}}$ of
${\cal B}=(3.4\pm0.3\pm0.6\pm0.4)\times10^{-4}$ and via a 
$\mathrm{D_0^{*0}}$ of
${\cal B}=(6.1\pm0.6\pm0.9\pm1.6)\times10^{-4}$, dominating this decay.

For the decay channel into $\DSP\PM\PM$, two broad resonances can
contribute to the decay in addition to the narrow resonance
$\mathrm{D_1^0}$. One is the newly-observed $\mathrm{D_0^{*0}}$.
In the fit to the Dalitz plot, its mass and width has been set to the
values obtained in the fit to $\mathrm{B^-\to D^+\pi^-\pi^-}$,
described above.
The second broad resonance contributing was identified with the
$\mathrm{D_1^{'0}}$. Its mass has been measured to be
$\mathrm{M_{D_1^{'0}}}=(2427\pm26\pm20\pm15)~\mathrm{MeV/c^2}$.
This is the first observation of this resonance.

The fit also gives the product branching ratios of the final state 
$\mathrm{B^-\to D^{*+}\pi^-\pi^-}$ decaying via 
$\mathrm{D_1^{'0}}$,
${\cal B}=(6.8\pm0.7\pm1.3\pm0.3)\times10^{-4}$, via a
$\mathrm{D_2^{*0}}$,
${\cal B}=(1.8\pm0.3\pm0.3\pm0.2)\times10^{-4}$ and via the newly
observed $\mathrm{D_1^{'0}}$,
${\cal B}=(5.0\pm0.4\pm1.0\pm0.4)\times10^{-4}$.

Though the new particles give a large contribution to the total decay
width in the decay modes used here, their large widths made it
difficult to observe them.

\subsection{D Mesons: $\mathrm{D_{sJ}}$}
\label{sec:1.4}
BaBar and CLEO observed two narrow resonances in the mass spectrum of
$\mathrm{D_s^{(*)+}}\PZ$. Their masses have been determined to 2317 \MeVc\ and
2460 \MeVc, respectively. In the following they will be referred to as
\DsJ\ and \DssJ.

Belle has confirmed both peaks in \epem\ annihilation, using the above
mentioned decay channels, and determined their masses to 
$M_\DsJ=(2317.2\pm0.5\pm0.9)~\MeVc$ and for the other state
$M_\DssJ=(2456.5\pm1.3\pm1.1)~\MeVc$. No information about their
quantum numbers was available so far.

In order to determine quantum numbers and branching ratios, the
additional constraints in decays of $\mathrm{B}$ mesons are
indispensable. An angular analysis of $\mathrm{B\to \overline{D}\Dj}$  
can help determine branching ratios and will reveal the quantum
numbers. 

In a sample containing $123\times 10^6$ \BBbar\ events, Belle searched
for the decays of B mesons to the new states \Dj\ accompanied by a
$\mathrm{\overline{D}}$
meson. Consistent signal yields are found from fits to the
distributions of \dE, \Mbc\ and
in the invariant mass of the $\mathrm{D_s}^{(*)}$ and the \PZ, the
significances are above $6\sigma$.

A clear signal ($7.4\sigma$) in another channel was found,
$\DssJ\to \mathrm{D}_s\gamma$, with a mass of
$\mathrm{M}_\DssJ=(2458.8\pm2.7\pm2.0)~\MeVc$ consistent with the
other channel. The large signal in this channel already rules out the
$0^\pm$ assignment for the \DssJ.
In other channels no signal was found, table~\ref{tab:1} lists the
different decay modes of the \DsJ\ and \DssJ.

\begin{table}
\caption{The upper two rows describe results for the decay
$\mathrm{B\to \overline{D}\DsJ}$, the lower five describe the results for
$\mathrm{B\to \overline{D}\DssJ}$}
\label{tab:1}       % Give a unique label
\begin{tabular}[h]{l@{\hskip 0.5cm}l@{\hskip 0.5cm}l}
decay channel & $\mathcal{B},~10^{-4}$ & significance \\ \hline\noalign{\smallskip}
$\mathrm{\DsJ\to D_s \pi^0}$
   &  $8.5^{+2.1}_{-1.9}\pm2.6$  & $6.1\sigma$ \\
$\mathrm{\DsJ\to D_s^* \gamma}$
   &  $<5.8~(2.5^{+2.0}_{-1.8})$  & $1.8\sigma$ \\ \hline
$\mathrm{\DssJ\to D_s^* \pi^0}$
   &  $17.8^{+4.5}_{-3.9}\pm5.3$  & $6.4\sigma$ \\
$\mathrm{\DssJ\to D_s \gamma}$
   &  $6.7^{+1.3}_{-1.2}\pm2.0$  & $7.4\sigma$ \\
$\mathrm{\DssJ\to D_s^* \gamma}$
   &  $<5.6~(2.7^{+1.8}_{-1.5})$  & $2.1\sigma$ \\
$\mathrm{\DssJ\to D_s \pi^+\pi^-}$
   &  $<1.2$   \\
$\mathrm{\DssJ\to D_s \pi^0}$
   &  $<1.2$ \\ \hline
\end{tabular}
%\vspace*{5cm}  % with the correct table height
\end{table}

Since in the latter decay, the quantum numbers of all other particles
except for the \DssJ\ are well known, an angular analysis can be
performed to narrow down further possible spin and parity assignments
assignments to this particle. The helicity angle distribution of the
\DssJ\ strongly favours a $\sin^2 \theta_{hel}$ distribution, which
supports the $1^+$ assignment to this particle. Here, the helicity
angle $\theta_{hel}$ is defined 
as the angle between \DssJ\ momentum in the $B$ meson rest frame and
the $\mathrm{D_s}$ momentum in the \DssJ\ rest frame. 

Belle also searched in \epem\ annihilation for these particles. 
The existence of the decay modes $\Dj\to\mathrm{D_s^{(*)}}\gamma$ is
confirmed, with a signal of $152\pm18$ events.
The ratio ${\cal
B}(\DssJ\to\mathrm{D_s}\gamma)$ over ${\cal
B}(\DssJ\to\mathrm{D_s^*}\pi^0)$ has been determined to
$0.63\pm0.15\pm0.15$, to be compared to $0.38\pm0.11\pm0.04$ determined
in $B$ decays. For the \DsJ, no signal is seen in the $\gamma$ channel
and an upper limit on the branching ratio of this decay of less than
5\% of the branching ratio of $\DsJ\to\mathrm{D_s^{(*)+}}\PZ$ is set
at the 90\% confidence level. 

It is worthwhile to notice is that the masses as well as mass
differences between the $0^+$ and the $1^+$ states in the $c\bar{u}$
system and the $c\bar{s}$ system are almost degenerate, i.e.
within uncertainties the same.

\section{Charm Fragmentation}
\label{sec:2}
The fragmentation of heavy quarks differs from that of light quarks
due to their higher masses. Many different model predictions exist, the
model of Peterson et al. being one of the most well known.

The data sample for this analysis consists of 3.65~\fb\ taken 60 \MeV\
below and 25.53~\fb\ at the \Ys\ resonance. In the latter sample, a
momentum cut removes contributions from $B$ decays. For now, only the
fragmentation of $c$ into \DSP\ has been measured; the analysis will
be extended to include
other charmed hadrons in the future. The \DSP\ has been
reconstructed in the fully charged decay channel $\DSP\to\DZ\PP$,
$\DZ\to\KM\PP$. Various binnings in the scaled momentum of the \DSP\
candidate have been used, 10 bins for comparison with other
measurements, 50 bins for the comparison between different models and
100 bins for the determination of the mean scaled momentum. The scaled
momentum \xP\ is defined as the momentum of the \DSP\ candidate in the
\Ys\ rest frame, scaled to its maximum allowed value of about $4.89~
\GeVc$.

This determination has been compared to two other measurements at
ARGUS and CLEO, which were based on a much smaller data set about 1/30
the size of the present analysis. The data agree very well.

The data has been compared to various models: the above-mentioned Peterson
et al, Kartvelishvili et al, Bowler, Collins and Spiller and the Lund
model. Table~\ref{tab:3} lists the value of the model parameters in
the minimum. The Bowler models agrees best with the data.

\begin{table}
\caption{\label{ffunctions}
Different fragmentation functions compared to the data. The free parameters
of each \frf s has been tuned for best agreement with data, regardless
of their influence on other distributions.}
\label{tab:3}       % Give a unique label
  \begin{tabular}{lcccc}
     \hline
     fragment'n fct &
       $\chi^2_{\mathrm{min}}$ & value at minimum & sample \\
     \hline
%     \hline
%
     Bowler et al. & 59.2/49 &
     $a=0.23$, $b=0.568$ & off-res \\
       & 166.8/24 & $a=0.21$, $b=0.532$ & on-res \\ % \hline
     Lund & 110.6/49 &
     $a=0.68$, $b=0.58$ & off-res \\
       & 517.0/24 & $a=0.61$, $b=0.58$ & on-res \\ % \hline
     Kartvelishvili & 122.2/49 &
     $\alpha_{c}=3.6$ & off-res \\
       et al. & 1050.7/24 & $\alpha_{c}=3.5$ & on-res \\ % \hline
     Collins-Spiller & 185.44/49 &
     $\varepsilon_{c}=0.089$ & off-res \\
       & 736.9/24 & $\varepsilon_{c}=0.093$ & on-res \\ % \hline
     Peterson et al. & 300.6/49 &
     $\varepsilon_{c}=0.054$ & off-res \\
       & 1613.1/24 & $\varepsilon_{c}=0.052$ & on-res \\ % \hline
     \hline
   \end{tabular}
%\vspace*{5cm}  % with the correct table height
\end{table}

The mean value for the scaled momentum of the \DS\ has been determined
with the off-resonance sample 
to $\langle \mathrm{x_P}(\DSpm) \rangle =
 0.610\pm0.003 \mathrm{(stat.)} \pm0.004 \mathrm{(syst.)}.$

\section{Conclusion}
Although being a dedicated $\mathrm{B}$ factory, the Belle detector at
the KEKB storage ring has been proven to give valuable contribution to
the charm physics sector. 
In Belle, the \ecS\ has been seen in both 
$\mathrm{B}$ decays and double \ccbar\ production.
The broad P-wave $\mathrm{D^{**}}$ states have been observed for the
first time.
The new particles observed by BaBar and CLEO have been confirmed, and
their possible quantum numbers have been been narrowed.
Last, but not least, the charm quark fragmentation into \DSP\ has been
measured with a much larger data sample.


\end{document}